\documentstyle[11pt,epsfig,epsf]{article}
\newcommand{\ba}{\begin{array}}
\newcommand{\ea}{\end{array}}
\newcommand{\bd}{\begin{displaymath}}
\newcommand{\ed}{\end{displaymath}}
\newcommand{\be}{\begin{equation}}
\newcommand{\ee}{\end{equation}}
\newcommand{\bea}{\begin{eqnarray}}
\newcommand{\eea}{\end{eqnarray}}

\def\q2 {q^2}

\begin{document}
\begin{flushright}
{\large 4th.~~~~~ ICPAQGP \\
26~-~30 November, 2001}  
\end{flushright}
\vskip1cm

\begin{center}
{\Large \bf Symmetry structure and phase transitions }\\[15mm]
{\bf Ashok Goyal \footnote{E-mail: agoyal@ducos.ernet.in}, Meenu Dahiya and Deepak Chandra }\\
{\em  Department of Physics and Astrophysics,}\\
{\em  University of Delhi, Delhi-110007, India.}\\[7mm]
\end{center}
\vskip 17pt
\renewcommand{\today}{}
\begin{abstract}
We study chiral symmetry structure at finite density and temperature in the presence of external magnetic field and gravity, a situation relevant in the early Universe and in the core of compact stars. We then investigate the dynamical evolution of phase transition in the expanding early Universe and possible formation of quark nuggets and their survival.
\end{abstract}
\begin{section}{Introduction}
Spontaneous symmetry breaking is one of the most important concepts of all unified gauge theories. The idea that underlying symmetries of nature are larger than that of the vacuum play a crucial role in the unification of forces. Of particular interest is the expectastion that at high temperatures, symmetries that are spontaneously broken today are restored and that during evolution, the Universe passed through a series of phase transitions from a higher symmetric phase to a lower symmetric phase associated with the spontaneous breakdown of gauge or perhaps global symmetry. In particular at $t=10^{-10}$s when the temperature was $\sim 200 GeV$, the Universe passed through the EWS breaking phase transition and later at $t=10^{-5}$s at $T=200 MeV$, there must have been a QCD phase transition from QGP to confined hadronic matter and also to chiral symmetry breaking phase transition. Since the vacuum structure of SSB theories is very rich, topologically stable configurations of gauge and Higgs fields in the form of domain walls, cosmic strings and monopoles on the one hand and non-topological solitons like Q balls, quark nuggets, solitone stars on the other may exist and may have observable signature. It is thus very instructive to investigate how the phase transition takes place in QFT in the environment of the early Universe and in the core of neutron stars where temperature, density, external electromagnetic field and external gravity may all play important role.\\
In GUTS Higgs play a most important role. They are fundamental scalars which give masses to fermions and gauge bosons through their VEVs. In theories like Technicolour models, Higgs are composite of some fundamental fermion fields. QCD is an example of QFT which is invarient under chiral transformations at the Lagrangian level in the absence of quark mass matrix, the dynamics of QCD are expected to be such that chiral symmetry is dynamically broken with the vacuum state acquiring a non-zero quark-antiquark condensate $<\bar qq>$, and the Goldstone theorem then requires the existence of approximately massless pseudoscalar mesons. To study chiral phase transition in QCD we need a non-perturbative treatment and a particularly attractive frame work to study is the Nambu-Jona-Lasino model. The linear Sigma model is another such model which has the advantage of being a renormalizable model in contrast to the NJL model which in 4-D is known to be non-renormalizable. An elegent and efficient way to study symmetry properties of vacuum at finite temperature and density in external envirornment is through the `Effective Potential' approach discussed extensively in the litrature. We compute here, in the one loop approximation, the effective potential in the presence of external magnetic field and gravity at finite temperature and density in the frame work of linear sigma and NJL model respectively. We then study the dynamic evolution of phase transition through bubble nucleation of hadronic phase. 
\end{section}
\begin{section}{Chiral symmetry in external magnetic field}
It has also been suggested \cite{linde1} that systems with
spontaneously broken symmetries may make a transition from broken
symmetric to restored symmetric phase in the presence of external
fields. Large magnetic fields with  strength upto $10^{18}$ gauss 
 have been conceived to exist \cite{ternov} at the time of supernova collapse
inside neutron stars and in other astrophysical compact objects  and
in the early Universe. 
Effect of such a strong magnetic field on 
chiral phase transition is  thus of great interest for baryon free
quark matter in the early universe and for high density baryon matter
in the core of neutron stars. To study chiral phase transition in QCD
 we need a nonperturbative treatment.
 Lattice techniques and the Schwinger-Dyson equations provide
specially powerful methods to study the chiral structure of QCD.
A particularly attractive frame work to study such systems is the 
linear sigma model originally proposed as a model for strong nuclear 
interactions. We will consider this as an effective model
for low energy phase of QCD and will examine the chiral symmetry properties 
at finite density and in the presence of external magnetic field. To fix 
ideas we consider a two flavor $SU(2) \times SU(2)$
chiral quark model given by the lagrangian.
\begin{equation}
{\mathcal L} = i \bar{\psi} \gamma^\mu \partial_\mu \psi
- g \bar{\psi}( \sigma+ i \gamma_5 \vec{\tau} . \vec{\pi}) \psi
+ \frac{1}{2} (\partial_\mu \sigma)^2
+ \frac{1}{2} (\partial_\mu \vec{\pi})^2 - U(\sigma,\vec{\pi})
\end{equation} 
where $\psi$  is the quark field $\sigma$ and $\vec{\pi}$ are the set of four scalar
fields and g is the quark meson coupling constant. The potential
 U$(\sigma ,\vec{\pi})$ is
given by \  
\begin{equation}
U( {\sigma}, \vec{\pi})= - \frac{1}{2} \mu^2 ( \sigma^2+ \vec{\pi}^2)+\frac{1}
{4}{\lambda} ( {\sigma}^2+ \vec{\pi}^2)^2
\end{equation}
For ${\mu}^2>0$ chiral symmetry is spontaneously broken. The $\sigma$ field
can be used to represent the quark condensate, the order parameter for chiral 
phase transition and the pions are  the Goldstone bosons. At the
tree level the sigma , pion and the quark masses are given by
\begin{equation}
m_{\sigma}^2=3{\lambda} {\sigma}_{cl}^2-{\mu}^2    ;    m_{\pi}^2={\lambda} {\sigma}_{cl}^2-{\mu}^2   ;    m_{\psi}^2 = g {\sigma}_{cl}
\end{equation}
 where ${\sigma}_{cl}^2=\frac{\mu^2}{\lambda} = f_{\pi}^2 $.
 We compute here, in the one loop approximation, the
effective potential in the presence of external magnetic field which is 
defined through an effective action $\Gamma(\sigma,B)$
 which is the generating functional of the one particle irreducible
 graphs. The effective potential is then given by 
\begin{equation}
V_{eff}(\sigma,B)=V_{0}(\sigma)+V_{1}(\sigma,B)
\end{equation}
where $V_{1}(\sigma,B)$ is obtained from the propagator function $G(\sigma,
B)$ by the usual relation
 $V_{1}(\sigma,B)=-\frac{1}{2 i} Tr \log G(\sigma,B)$.
 Alternatively one can compute the shift
in the vacuum energy density due to zero-point oscillations of the fields 
considered as an ensemble of harmonic oscillators \cite{kirshnitz}. We thus
require energy eigenvalues(excitations) of particles in the magnetic field,
which can be easily obtained, and in the absence of anomalous magnetic moment
for uniform static magnetic field in the z-direction for a particle of mass
M,charge q and spin J, are given by  
\begin{equation}
\label{one}
E(k_{z},n,J_{z})={(k_{z}^2+M^2+(2 n+1-sign(q)\,\,j_{z}) \left\vert 
q \right\vert B)}^2
\end{equation}
where n represents the landau level.
In the presence of magnetic field, all we need to do is to replace the phase
space integral
 $\int \frac{d^4 k_{e}}{(2 \pi)^4} $  by  $\frac{e B}{2 \pi}
 \sum_{n=0}^\infty \frac{d^2 k_{e}}{(2 \pi)^2}$   
and the energy by expression (\ref{one})  for charged particles.
The one loop effectiv potential can now be calculated by standard techniques by dimensional regularization. The effective potential has poles which can be absorbed in counter terms. The finite part depends on
the exact renormalization conditions that are imposed. In what follows
we would use the $\overline{MS}$ renormalization scheme. 
 To proceed further we first consider the 
case $\frac{M^2}{2 e B}<1$ , keeping leading terms and adding the contributions of all the charged particles
the total $ V_{eff}(\sigma,B)$ for the sigma model at the one loop
level is thus given  by \cite{goyal1}
\begin{eqnarray}
\label{seven}
V_{eff}(\sigma,B) &=& -\frac{1}{2} \mu^2 \sigma^2+\frac{\lambda}{4} \sigma^4
\nonumber \\
&+&\frac{1}{64 \pi^2}\,\, (3 \lambda \sigma^2-\mu^2)^2\,\, \log\,\,
(\frac{3 \lambda\sigma^2-\mu^2}{m_\sigma^2}-\frac{3}{2}) \nonumber \\
&+& \frac{1}{64 \pi^2}\,\, (\lambda\sigma^2-\mu^2)^2\,\, \log\,\, (\frac{\lambda \sigma^2
-\mu^2}{m^2}-\frac{3}{2})
\nonumber \\
&-&\frac{e B}{16 \pi^{2}}\,\, (\lambda \sigma^2-\mu^2)\,\, \log 2
\nonumber \\
&-& \frac{N_{c}}{16 \pi^2} \sum_{flav}\,\, [g^{4} \sigma^{4}\,\,
(\,\, \log \frac{g^{2}\sigma^2}{m_{f}^2}-\frac{3}{2})
\nonumber \\
&+&\frac{2}{3}\,\, (\left\vert q \right\vert B)^{2}\,\, \log \frac{g^{2} \sigma^{2}}{
m_{f}^2}]
\end{eqnarray}
For $\frac{\left\vert q \right\vert B}{M^2} > 1$  ,the last term in eqn.(\ref{seven}) is replaced by  $\frac{\left\vert q \right\vert BM^2}{8\pi^2}(1-logM^2)$ summed
over flavors.
In figure 1, we plot $V_{eff}(\sigma,B)$ as a function of $\sigma$ for 
different values of magnetic field and compare it with the case of zero 
magnetic field by. As input 
parameters we choose the constituent quark mass $m_f$= 500 MeV, sigma mass
$m_\sigma$=1.2 GeV and $f_{\pi}$=93 MeV.
We find that in the presence of intense magnetic fields the chiral symmetry breaking is enhanced. For magnetic field large compared to $m_f^2$,  we observe that though the fermionic contribution is towards symmetry
restoration, it is not enough to offset the contribution of charged goldstone pions.
In order to study chiral symmetry restoration in the case of neutron stars 
as a function of chemical potential $\mu$ associated with finite baryon
 number density we employ the imaginary time formalism by summing over
 Matsubara frequencies. This amounts to adding the fermionic free energy to
 the one loop effective potential and is given by 
\begin{equation}
V_{1}^{\beta}(\sigma)=-\frac{\gamma}{\beta} \int \frac{d^3 k}{(2 \pi)^3}
\,\, \ln \,\,(1+e^{-\beta(E-\mu)})
\end{equation}
 which in the presence of static uniform magnetic field becomes 
\begin{equation}
V_{1}^{\beta}(\sigma)=-\frac{\gamma}{\beta}\,\, \frac{e B}{2 \pi} \sum_{n=0}^{\infty}
\int_{0}^{\infty} \frac{d k_{z}}{2 \pi}\,\, \ln\,\, (1+ e^{-\beta (E-\mu)})
\end{equation}
where $\gamma$ is the degeneracy factor and is equal to 2$N_c$ for each quark flavor.
We consider cold dense isospin symmetric quark matter for which the integrals can be
performed analytically.
The baryon number density corresponding to the chemical potential $\mu$ is given by
the usual thermodynamical relations.
\begin{equation}
N_{B}(\mu,0)=\frac{1}{3} \sum_{flav}\,\, \frac{\gamma}{6 \pi^{2}} \,\, (\mu^{2}-g^{2} \sigma^2)^{\frac{3}{2}}
\end{equation}
 and
\begin{equation}
N_{B}(\mu,B)= \frac{1}{3} \sum_{n=0}^{n_{max}}\,\,\frac{\gamma \left\vert
 q \right\vert B}{4 \pi^2}\,\, (2-\delta_{\mu,0})\,\, \sqrt{\mu^{2}-g^{2} 
\sigma^{2}-2 n \left\vert q \right\vert B}
\end{equation} 
for zero and finite magnetic field respectively.
Here $n_{max}$= Int $\left\vert \frac{\mu^2-g^2 \sigma^2}{2 \left\vert q
\right\vert B} \right\vert$.
To study chiral symmetry behavior at finite density in the presence of uniform
magnetic field, we minimize effective potential with respect to the
order 
parameter $\sigma$ for fixed values of chemical potential and magnetic field ( which then fixes the baryon density ). The results are shown in
figure 2 
where we have plotted the order parameter $\sigma$ as a function of
density 
at T=0 for different values of magnetic field. The solution indicates a first
order phase transition. The actual transition takes place at the point where
the two minima of the effective potential at $\sigma$=0 and 
$\sigma$=$\sigma$($\mu$,B) non zero become degenerate. The lower values of $\sigma$ (shown by dotted curves) are unphysical
in the sense that they do not correspond to the lowest state of energy. We find that magnetic field continues to enhance
chiral symmetry breaking at low densities as expected but as the magnetic field
is raised the chiral symmetry is restored at a much lower density compared to
the free field finite density case. This can be clearly seen from figure 3 where we have plotted the phase diagram in terms of baryon density and magnetic field.
\end{section}        

\begin{section}{Chiral symmetry in NJL model in curved space time}

The NJL model four fermion theory in flat space-time in arbitrary dimensions
has been studied \cite{inagaki1} using the 1/N expansion method. It is shown that chiral 
symmetry is restored in the theory under consideration for sufficiently high 
temperature or chemical potential. It is found that for space-time dimensions
$2 \leq D < 3$ both a first order and second order phase transition occur
 depending on the value of the four fermion coupling while 
for $3 \leq D < 4$  only the second order phase transition exists.
Here we study the chiral phase structure of the four fermion theory in curved space-time at finite temperature and density. The Nambu Jona Lasinio model in curved space time is defined by the action
\begin{equation}
S=\int d^D x \, \sqrt{(-g)} \,[\,\, i \,\bar{\psi} \gamma^\mu(x) \bigtriangledown_\mu \psi
+\frac{\lambda}{2 N} (\,\,{(\bar{\psi}\psi)^2+(\bar{\psi}i\, \gamma_5 
\,\vec{\tau} \,\psi)^2})\,\,]
\end{equation}
where $g$  is the determinant of the space time metric ,$\gamma^\mu(x)$ the
Dirac matrix in curved space-time, $\bigtriangledown_\mu \psi$ the covariant 
derivative of the fermion field $\psi$ and N is the number of colours, we take the number of flavors to be two.
We work in the scheme of the 1/N expansion and perform our calculations in the
leading order of the expansion.
For practical purposes it is more convenient to introduce the auxiliary
fields $\sigma$ and $\vec{\pi}$ and consider the equivalent action.
\begin{equation}
S=\int d^D x \sqrt{(-g)}\, [ i\, \bar{\psi} \gamma^\mu(x) 
\bigtriangledown_\mu \psi-\frac{ N}{2 \lambda} (\sigma^2+\pi^2)
- \bar{\psi}( \sigma+ i \,\gamma_5 \,\vec{\tau} \, \vec{\pi}) \psi ]
\end{equation}
Replacing $\sigma$ and $\vec{\pi}$  by the solutions of the Euler-Lagrangian 
equations arising from (12) we reproduce the action (11). This is because the fields 
$\sigma$ and $\vec{\pi}$ are not independent degrees of freedom in (12) and the
Euler-Lagrangion equations for  $\sigma$ and $\vec{\pi}$ are infact constraint equations which fix $\sigma$ and $\vec{\pi}$ given $\psi$ and $\bar \psi$.
If a non vanishing vacuum expectation value is assigned to the auxiliary
field $\sigma$, then there appears a mass term for the fermion field
$\psi$ and the discrete chiral symmetry is eventually broken.
The effective potential (with N factored out) in the leading order 
of the 1/N expansion is then given by 
\begin{equation}
V(\sigma, \pi)= \frac{1}{2 \lambda} (\sigma^{2}+\pi^{2}) + i \,Tr \,  ln \,
S\,(x,x;s) \,\Bigg\vert_{s=\sigma+\,i \gamma_5\, \vec{\pi}}
\end{equation}
In equation (13) the variables $\sigma$ and $\vec{\pi}$ are regarded as constantand the Green function S is the solution of the equation
\begin{equation}
(i \gamma^\mu(x) \bigtriangledown_\mu -s) S(x,y;s)= \frac{1}{\sqrt{(-g(x))}} \delta^D(x-y)
\end{equation}
Thus the effective potential is described by the two point Green's function 
S(x,x;s) of the massive free fermion in curved space-time.
Using the Green function obtained in the approximation of keeping only 
linear terms
in the curvature, the 
effective potential which in D-dimensions read as follows:
\begin{eqnarray}
V( \sigma, 0) &=& \frac{\sigma^{2}}{2 \lambda} -i Tr  \int_{0}^{\sigma} d s \int
 \frac{d^D k}{(2 \pi)^4} [ (\gamma^{a}  k_{a}+s) \frac{1}{k^2-s^2}
\nonumber \\
&&-\frac{R}{12} (\gamma^{a}  k_{a}+s) \frac{1}{(k^2-s^2)^{2}}
+ \frac{2}{3} R_{\mu,\nu} k^{\mu} k^{\nu} (\gamma^{a}  k_{a}+s) 
\nonumber \\
&&\times \frac{1}{(k^2-s^2)^{3}} - \frac{1}{2} \gamma^{a} J^{c d} R_{c d a \mu} 
k^{\mu} \frac{1}{(k^2-s^2)^{2}}]
\end{eqnarray} 
 where $ J^{a\, b}= \frac{1}{4} [ \gamma^{a} , 
\gamma^{b} ] $ and latin indices are  vierbein indices.
The effective potential is divergent in two and four dimensions and is
finite in three dimensions in the leading 1/N expansion. The four fermion
theory is renormalizable in 2-D flat space.
Using the renormalization condition $ \frac{\partial^{2} V_{0} (\sigma)}{\partial\sigma^{2}}
 \Bigg\vert_{\sigma=M} = \frac{M^{D-2}}{\lambda_{r}}$
where M is the renormalization scale.
The renormalised effective potential in D dimensions is given by 
\begin{eqnarray}
\frac{ V(\sigma,0)}{M^{D}} &=& \frac{1}{2 \lambda_{r}} \frac{\sigma^{2}}{M^{2}}
+\frac{Tr \mathbf {1}}{ 2 \,(4 \pi)^{\frac{D}{2}}} \,\,(D-1)\, \Gamma\,(1-\frac{D}{2}) \,\,
\frac{\sigma^{2}}{M^{2}}
\nonumber \\
&& - \frac{Tr \mathbf{1}}{ (4 \pi)^{\frac{D}{2}} D} \,
\Gamma\,(1-\frac{D}{2}) \,\,\frac{\sigma^{D}}{M^{D}}-
 \frac{Tr \mathbf{1}}{(4 \pi)^{\frac{D}{2}}} \frac{R}{M^2} \frac{1}{ 24} 
\nonumber \\
&&\times \Gamma\,(1-\frac{D}{2}) \,\,\frac{\sigma^{D-2}}{M^{D-2}} 
\end{eqnarray}
We now obtain the four dimensional limit of the NJL model in the MS 
renormalization scheme given by
\begin{eqnarray}
\frac{V (\sigma,0)}{M^{4}} &=& \frac{1}{2 \lambda} \,\,(\frac{\sigma}{ M})^{2}
 -\frac{1}{4 \,\pi^{2}}\,\, ( 1+3\,\, ln \,4 \pi-3 \, \gamma) \,\,
(\frac{\sigma}{M})^{2}
\nonumber \\
&& -\frac{1}{8 \,\pi^{2}}\,\, (\,ln\,\, (\frac{ \sigma}{M})^{2}-\frac{3}{2}-ln \,4 \pi+\gamma) \,\, (\frac{\sigma}{M})^{4}
\nonumber \\
&& -\frac{R}{48 \,M^{2} \pi^{2}} \,\,(\,\,ln \,(\frac{ \sigma}{M})^{2} -1
 -ln\,\, 4 \pi+\gamma)\,\,(\frac{\sigma}{M})^{2}
\end{eqnarray}
Alternatively one could  regularize the divergent part by cutting off the momentum
integral at finite cutoff $\Lambda$ \cite{inagaki2}.
This gives
\begin{eqnarray}
V (\sigma,0) &=& \frac{\sigma^{2}}{2 \lambda} -\frac{1}{(4 \pi)^{2}}\,\,
 [\,\, \sigma^{2} 
\Lambda^{2} + \Lambda^{4}\, ln\, (1+\frac{\sigma^{2}}{\Lambda^{2}}) -\sigma^{4}\,
 ln\, (1+ \frac{\Lambda^{2}}{\sigma^{2}})]
\nonumber \\
&&-\frac{1}{ (4 \pi)^{2}}\, \frac{R}{6} \,\, [\,\, -\sigma^{2} \, 
ln \, (1+ \frac{\Lambda^{2}}{\sigma^{2}})
+ \frac{\Lambda^{2}\sigma^{2}}{\Lambda^{2} +\sigma^{2}}\,\,]
\end{eqnarray}
The two expression can be shown to be equivalent after carrying out the 
renormalization of the coupling constant.
                     The ground state of the theory is determined by the minimum
of the effective potential (17) namely, by solving the gap equation
\begin{equation}
\frac{\partial V(\sigma,0)}{\partial \sigma}\big\vert_{\sigma=m}=0
\end{equation}
For $ \lambda_{r} >\lambda_{c}=\frac{2 \pi^{2}}{1 + 3 ln \,4 \,\pi-3\,\gamma}$,
the minimum of the effective potential is located at the non-vanishing $\sigma$,the chiral symmetry is broken down dynamically and a dynamical fermion mass is
generated. At the critical point the effective potential has two degenerate
local minima obtained by putting
\begin{equation}
V(\sigma,0)=V(m,0)=0
\end{equation}
The solution of the gap equation and the value of the
 critical curvature  can be obtained numerically.
First, we fix the coupling constant $\lambda_{r}$ greater than the critical value $\lambda_{c}$ corresponding to the  broken symmetric phase.
To study the phase structure in curved space-time we evaluate the effective 
potential  (17) by varying the space-time curvature. We see from Fig.4
 that the chiral symmetry is restored  as R is increased 
for a fixed $\lambda$. The phase transition induced by curvature is of 
first order. 
The effective potential $V^{\beta}(\sigma,0)$ at finite temperature, density and curvature can be obtained by standard techniques as in Section 2 and is given by {\cite{goyal2}}
\begin{eqnarray}
V^{\beta} (\sigma,0) &=& \frac{\sigma^{2}}{2 \,\,\lambda} -\frac{1}{4 \pi^{2}}
\,\, (\, 1+3 \,ln \, 4 \pi- 3 \,\gamma)\,\, \sigma^{2}
\nonumber \\
&& -\frac{1}{8 \pi^{2}} (\,ln \frac{ \sigma^{2}}{M^{2}}-\frac{3}{2}
-ln \,4 \pi+\gamma \,)  \sigma^{4}
\nonumber \\
&& -\frac{R}{48 \pi^{2}} (\, ln \,\frac{ \sigma^{2}}{M^{2}} -1 
-ln \,4 \pi+\gamma) \sigma^{2}
\nonumber \\
&&  \frac{-2}{ \beta \pi^{2}} \int k^{2} d k
 [ln (1+e^{- \beta [\sqrt{ {(k^{2}+\sigma^{2})}}- \mu] } +\mu \rightarrow -\mu]
\nonumber \\
&& +\frac{R \sigma^{2 }}{12 \pi^{2}} \int \frac{k^{2} d k}
{(k^{2}+\sigma^{2})^{ \frac{3}{2}}} ( \frac{1}{ 1+e^{\beta 
[ \sqrt{ (k^{2} + \sigma^{2})} -\mu ]}}+ \mu \leftrightarrow -\mu)
\nonumber \\
&&+\frac{R \sigma^{2 } \beta }{6 \pi^{2}} \int \frac{k^{2} d k}
{(k^{2}+\sigma^{2})} (  \frac{ e^{\beta[ \sqrt {( k^{2}+\sigma^{2 }}-\mu)]}}
{ ( 1+e^{ \beta [ \sqrt{ (k^{2}+\sigma^{2})}- \mu]})^{2} } \mu \leftrightarrow -\mu)
\end{eqnarray} 
At high densities and low temperatures or at high temperatures and low densities,the integrals in (21) can be carried out analytically by the standard techniques. 
In Fig.5 we have shown the phase boundary in temperature, chemical potential 
plane for different values of curvature. We notice that with
 the increase in R, phase transition takes place at lower temperature and
density. In Fig.6 we have shown the phase boundary curve in the temperature,
curvature and chemical potential plane.  
\end{section}
\begin{section}{Dynamical evolution of quark-hadron phase transition through bubble nucleation}
\indent It is well known that a phase transition from quark gluon plasma to 
confined hadronic matter must have occurred at some point in the evolution of 
the early Universe, typically at around $10-50 \mu s$ after the Big Bang. This
leads to an exciting possibility of the formation of quark nuggets through 
the cosmic separation of phases {\cite{wit}}. As the temperature of the 
Universe falls below the critical temperature $T_c$ of the phase
transition, the quark gluon plasma super cools and the transition proceeds
through the bubble nucleation of the hadron phase. The typical
distance between the nucleated bubbles introduces a new distance scale to the
Universe which depends critically on the super cooling that takes place. 
As the hadronic bubbles expand, they heat the surrounding plasma, shutting off
further nucleation and the two phases coexist in thermal equilibrium. The
hadron phase expands driving the deconfined quark phase into small regions of
space and it may happen that the process stops after the quarks reach 
sufficiently high density to provide enough pressure to balance the surface
tension and the pressure of the hadron phase. The quark matter trapped in 
these regions constitute the quark nuggets. The number of particles trapped in 
the quark nugget, its size and formation time are dependent sensitively on the
degree of super cooling. The duration of the phase transition also depends on 
the expansion of the Universe and on other parameters like the bag pressure $B$
and the surface tension $\sigma$. 
\par
The quark nuggets formed in the small super cooling scenario are in a hot
environment around the critical temperature $T_c$ and are susceptible
to evaporation from the surface {\cite{Alcock} } and to boiling
through subsequent  hadronic bubble nucleation inside the nuggets 
{\cite{Olinto}}. However in the large super cooling scenario we
have the interesting possibility of these nuggets forming at
a much lower temperature than $T_c$ due to the long duration of the
transition and consequent expansion of the Universe. Alcock and Farhi 
{\cite{Alcock}} have shown that the quark nuggets with baryon numbers 
$\leq 10^{52}$ and mass $\leq 10^{-5} M_{\odot}$ are unlikely to survive 
evaporation of hadrons from the surface. Boiling was shown to be even
more efficient mechanism of nugget destruction 
{\cite{Olinto}}. These
results were somewhat modified by Madsen et.al. {\cite{Madsen}} by
taking into account the flavor equilibrium near the nugget surface
for the case of evaporation and by considering the effect of
interactions in the hadronic gas for the case of boiling. In
the large super cooling scenario the time of formation of these nuggets
can be quite late when the number of baryons in the horizon 
(of size $\sim 2t$) is large and temperature much lower. These nuggets
can easily survive till the present epoch.
\par

There have been recent observations by gravitational micro lensing 
{\cite{Alcock1}} of dark objects in our galactic halo having masses of
about $0.01-1$ solar mass. If these objects have to be identified with
quark nuggets, they could only have been formed at a time later than
the time ($\sim 50-100 \mu s$) when the Universe cooled through $T_c$.
S
 When the early Universe as a quark-gluon 
thermodynamic system cools
through the critical temperature $T_c$, energetically, the new phase 
remains unfavorable as there is 
free energy associated with surface of separation between the phases. Small
volumes of new phase are thus unfavorable and all nucleated bubbles with
radii less than critical radius collapse and die out. But those with radii 
greater than the critical radius expand until they coalesce with each other. 
So super cooling occurs before the new phase actually appears and is then 
followed by reheating due to release of latent heat. 
The bubble nucleation rate {\cite{Fuller}} at temperature T is given by
\begin{equation}
I=I_o e^{-\frac{W_c}{T}} 
\end{equation}
where $I_o$ is the prefactor having dimension of $T^4$. The prefactor used 
traditionally in early Universe studies {\cite{Fuller}} is given by 
$I_o=(\frac{W_c}{2\pi T})^{\frac{3}{2}}T^4$. Csernai and Kapusta
{\cite{Csernai}} have recently computed this
prefactor in a coarse-grained effective field theory
approximation to QCD and give 
$I_o = \frac{16}{3\pi}
 (\frac{\sigma}{3T})^{\frac{3}{2}}\frac{\sigma \eta_q R_c}
{\xi^4_q (\Delta w)^2}$ 
 where $ \eta_q=14.4T^3 $  is the shear
 viscosity in the plasma phase, $\xi_q$ is a correlation length of 
 order 0.7 fm in the plasma phase and $ \Delta w$ is the difference
 in the enthalpy densities of the two phases. In this letter we use both
 these prefactors for comparison. The critical bubble radius $R_c$ and  
 the critical free energy $W_c$ are obtained by maximizing the thermodynamic
 work expended to create a bubble and are given by 
$R_c= \frac{2 \sigma}{P_h(T)-P_q(T)} $ and  
$W_c = 4\pi \sigma R^2_c (T)/3  = \frac{16\pi \sigma^3}{3\Delta P^2}$ 
where $\Delta P =P_h(T)-P_q(T)$  is the pressure difference in hadron
 and quark phase.
\par
For simplicity we describe the quark matter by a plasma of massless
$u$,$d$ quarks and massless gluons without interaction. The long range
non-perturbative effects are parameterized by the bag constant $B$. 
The  pressure in the QGP phase is given by $P_q(T)=\frac {1} {3} g_q \frac {\pi^2}
 {30} T^4-B$ where $g_q\sim 51.25$ is the effective number of degrees
of freedom. In the hadronic phase the pressure is
given by $P_h(T) =\frac{1}{3} g_h\frac{\pi^2}{30}T^4$ where $g_h\sim 17.25$,
taking the three pions as massless.
\par
The fraction of the Universe $h(t)$ which has been converted to hadronic 
phase at the time $t$ is
given by the kinetic equation
\begin{equation}
h(t)=\int_{t_c}^{t} I(T(t'))[1-h(t')]V(t',t)\bigg[\frac {R(t')} {R(t)}\bigg] ^3 dt'
\end{equation}
where $V(t',t)$ is the volume of a bubble at time $t$ which was
nucleated at an earlier time $t'$ and $R(t)$ is the scale factor.
This takes bubble growth into account
and can be given simply as
 \begin{equation}
V(t',t)=\frac{4\pi}{3}\Bigg[ R_c(T(t'))+\int_{t'}^{t} \frac{R(t)}
{R(t'')} v(T(t''))dt''\Bigg]^3
\end{equation}
where $v(T)$ is the speed of the growing bubble wall and can be taken
to be $v(T)=v_o [ 1- \frac{T}{T_c} ]^{\frac{3}{2}}$ where
$v_o=3c$. This has the correct behavior in that closer $T$ is to
$T_c$ slower do the bubbles grow. When $T=\frac{2}{3}T_c$ we have
$v(T)=\frac{1}{\sqrt{3}}$ the speed of sound of a 
massless gas. For $T<\frac{2}{3}T_c$ which occurs when there is
large super cooling, we use the value
$v(T)=\frac{1}{\sqrt{3}}$. 
\par
The other equation we need is the dynamical equation which couples time
evolution of temperature to the hadron fraction $h(t)$. We use the two
Einstein's equations as applied to the early Universe neglecting curvature.
\begin{equation}
\frac{\dot{R}}{R}=\sqrt{\frac{8\pi G} {3}} \rho^{\frac{1}{2}}
\end{equation}
\begin{equation}
\frac{\dot{R}}{R}=-\frac{1}{3w} \frac{d\rho}{dt} 
\end{equation}
where $w=\rho+P$ is the enthalpy density of the Universe at time $t$.
The energy density in the mixed
phase is given by $\rho(T)=h(t)\rho_h(T)+[1-h(t)]\rho_q(T)$, where
$\rho_h$ and $\rho_q$ are the energy densities in the two phases at
temperature $T$ and similarly for enthalpy. We numerically
integrate the coupled dynamical equations (23),(25) and(26) to study the
evolution of the phase transition starting above $T_c$ at some
temperature $T$ corresponding to time $t$
obtained by integrating the Einstein's equation (25) and (26) in the quark
phase.
The number density of nucleated sites at time $t$, is given by
\begin{equation}
N(t)=\int_{t_c}^{t} I(t')[1-h(t')]\bigg(\frac {R(t')} {R(t)}\bigg)^{3}dt'
\end{equation}
Therefore the typical separation between nucleation sites is 
 $l_n=N(t)^{-1/3}$.
This distance scale will eventually determine the number of quarks in
a nugget. This scale can be up to $10^{12}$ Km depending on the
parameters $B$ and $\sigma$ which
correspond to a distance of $\sim 1.4 Mpc$ today. The nuggets could
be points in space around which, later in time, the matter in the
Universe may have gravitationally clustered to give the observed large
scale structures in the Universe. The observable separation of 
galaxies in the Universe
can be a remnant of this transition with the centers of the galaxies
being the quark nuggets. Of course the collision of bubbles and their
random nucleation and interaction will also lead to clustering of the
nuggets, which can qualitatively explain the clustering of galaxies.
To get an idea about the super cooling before nucleation begins, we can
plot the nucleation time as a function of temperature, defined by 
$\tau^{-1}_{nucleation}=\frac{4\pi R_c^3}{3}I$. The quark
number density is given by 
$n_q=\frac{2}{\pi^2}\zeta (3)(\frac{n_q}{n_\gamma})T^3$ where $
\frac{n_q}{n_\gamma}$ is the quark to photon ratio estimated from the
abundance of luminous matter in the Universe to be roughly equal to 
$3\times 10^{-10}$. The quark nuggets have $N_q$ quarks at time
$t$ given by the number of quarks in a volume 
$\frac{1}{N(t)}$, i.e. $N_q=\frac{n_q}{N(t)}$. The nucleation sites
are actually randomly distributed, but we expect a distribution of
quark numbers around $N_q$. The average temperature at which nuggets 
are formed when bubbles
coalesce is obtained by finding the average time at which the expanding bubble
surfaces meet. Assuming a cubic lattice, we have done this numerically to
get the corresponding time $t_f$ and temperature $T_f$ for different 
values of $B$ and
$\sigma $. When the fraction of the space occupied by the bubbles is
around $50$ percent, we expect the bubbles to meet in an ideal
picture,i.e. if all bubbles are essentially nucleated at one instant
which is the maximum nucleation time and they all have the same
radius. However we have a distribution of expanding bubble sizes
because of the different points of time at which they were
nucleated. Therefore the estimate of the time of nugget formation by
treating all bubbles to be of the same size is an underestimate. We
find that hadron fraction $h(t)$ is only around $.12$ when bubbles
meet by this criteria. However we do not expect this to change
qualitatively the broad picture of the transition and the nugget
formation apart from reducing the formation time.
\par
 Fig.7 {\cite{goyal3}} shows the temperature as a function of time. It
is clear from this diagram that reheating takes place as nucleation
starts with the release of latent heat. As $\sigma$ increases, the
super cooling is larger and reheating is slower. The transition takes 
much longer to complete with more chance of nugget formation. This 
allows the nuggets to be formed at a much lower temperature when 
bubble walls meet. For low supercooling there is rapid reheating, temperature
reaches $T_c$ and the phase transition is completed very swiftly with
no chance of any nugget formation. Fig.8 shows the fraction $h(t)$ of the
Universe in hadron phase as a function of time. For small values of
$\sigma$ the transition completes quickly as $h(t)$ goes to $1$. But
for larger $\sigma$ it takes a larger time for $h(t)$ to become
$1$. We also notice that in the large super cooling scenario the Kapusta
prefactor becomes much bigger than the standard one by many orders of
magnitude. This makes the nucleation rate as well as the reheating
faster.In the case of low super cooling the two prefactors give 
almost identical results. The
number of quarks in the horizon $N_{qH}$ at time $t$ is 
$N_{qH}\sim n_q(\frac{4}{3}
\pi t^3 )\sim(\frac{n_q}{n_\gamma})\frac
{2\zeta(3)} {3\pi^2} T^34\pi t^3$
and we find that for all interesting cases $N_q \le N_{qH}$ and this number
is very sensitively dependent on the surface tension. Physically it is
possible to have $N_{qH}\ge N_q\ge 10^{52}$ for some values of the
parameters B and $\sigma$. In table I below we list some physical quantities 
for some representative values of $B$ and $\sigma$.

\[ {\begin{array}{cccccccc}
\hline
 B^{1/4} &\sigma & T_c & t_f & T_f & N_q & N_{qH} & l_n \\ 
 MeV  & MeV fm^{-2} & MeV & \mu s & MeV &   &  & m  \\
\hline
235 & 50 & 169 & 12.1 & 169 & 2.6\times 10^{28} & 7\times 10^{52}  & 8\times 10^{-3}	 \\
145 & 57.1 & 104.4 & 34.7 & 99.9 & 9.4\times 10^{35} & 3.4\times 10^{53} & 4.6\\
125 & 57.1 & 90 & 56.1 & 78.8 & 1.1\times 10^{39} & 7.3 \times 10^{53} & 63\\
125 & 77 & 90 & 1511.8 & 17.9 & 6.3\times 10^{48} & 1.6\times 10^{56} & 4.9\times 10^{5}\\
100 & 39.5 & 71.9 & 2595 & 13.9 & 1.2\times 10^{50} & 3.8\times 10^{56} & 8.4\times 10^{5}\\
113 & 57.1 & 81.3 & 5138 & 12.3 & 9\times 10^{49} & 2\times 10^{57} & 1.7\times 10^6\\	
\hline
\end{array}} \]
\end{section}
\begin{section}{Conclusions}
In conclusion we have examined the chiral symmetry behavior of
 the Linear Sigma model in the presence of static, uniform magnetic field
at the one loop level at zero density and at densities relevant in the
core of neutron stars. We find that the contribution of scalar and
fermion loops leads to an increase in chiral symmetry
breaking. At high densities too, this effect persists and for magnetic fields
of strength upto $10^{18}$ Gauss, there is enhancement in chiral symmetry 
breaking resulting in the restoration of symmetry at densities higher than if
no magnetic field were present. However, in the case of high
magnetic field $B\geq10^{19}$ Gauss the chiral symmetry is restored 
at lower densities. Thus in the core of neutron stars, if the
nuclear matter undergoes a transition to deconfined quark matter, the presence
of magnetic field would imply the existence of massive quark matter due to
enhancement in chiral symmetry breaking. This would affect the equation of 
state and will have astrophysical implications.
We then investigated the phase structure of NJL model in curved space time in the linear curvature approximation and found that in the presence of external gravity with positive R, the chiral symmetry restoration is first order even as in flat space-time, the transition is second order with temperature and becomes first order in the presence of density. We find that with the increase in R phase transition takes place at lower temperature and density.
Regarding evolution of quark-hadron phase transition detailed dynamics of the quark-hadron transition in the
early Universe show that the evolution of the Universe does not
necessarily follow the small super cooling scenario and certain choices
of B and $\sigma$ can have a bearing on the present state of the
Universe. As nuggets with $N_q\ge 10^{52}$ are expected to survive the
transition, they will contribute to the density of the Universe.  We have explored in detail the possibility of 
nugget formation and also estimated their average separation, time of
formation, quark content and survivability. Clearly,
the analysis can be improved by taking interactions into account in
both the phases and also bubble interactions may be incorporated in
the calculations, the qualitative results however are not expected to change. 
Thus if the nuggets
studied above are indeed formed in a much cooler environment, they
could contribute significantly to the missing mass in the Universe and
be candidates for dark matter. 

\end{section}

{\bf Acknowledgements}
A.G. would like to thank the organisers and participants for providing the opportunity to present this work and for making the conference stimulating and enjoyable.

\pagebreak

{\bf Figure Captions}
\vskip 1 cm
Figure 1. Effective Potential in units of $(100 MeV)^4$ as a function of
$\sigma(MeV)$ for different values of the magnetic field. The curves a, b, c, d and e are for B=0, $10^{16}, 10^{18}, 10^{19}$ and $3\times10^{19}$ Gauss respectively.    
\vskip 0.5 cm
Figure 2. Chiral condensate $\sigma(MeV)$ as a function of baryon density
in $f_m^{-3}$ for magnetic field B=0,$10^{16},10^{18} and 10^{19}$ Gauss respectively.
\vskip 0.5 cm
Figure 3. Phase diagram as a plot of magnetic field  versus baryon density
in units of $fm^{-3}$.
\vskip 0.5cm
Figure 4. Behaviour of Effective Potential $\frac{V}{M^{4}}$ as a function of $\frac{\sigma}{M}$ for different values of the curvature for $\lambda > 
\lambda_{cr} ( \lambda=10)$. The curves a, b, c, d, e and f are 
for R = 16, 13, 12, 8, 0 and -2 respectively.
\vskip 0.5cm
Figure 5. Temperature vs chemical pot. for different values of curvature. The curves a, b, c, d, e and f are for R/$M^{2}$ = 0, 2, 4, 6, 8 and 10 respectively.
\vskip 0.5cm
Figure 6. Phase diagram in the temperature, curvature and chemical potential plane.
\vskip 0.5cm
Figure 7. Temperature as a function of time. Solid and dashed curves are for
 $B^{1/4} =100 MeV$ and $\sigma =
39.5 MeV fm^{-2}$ with the standard and the Kapusta prefactors respectively.
Long dashed and dotted curves are for $B^{1/4} =113 MeV$ and $\sigma =
57.1 MeV fm^{-2}$ with the standard and the Kapusta prefactors respectively.
\vskip 0.5cm
Figure 8. The hadron fraction as a function of time. Curves as in fig. 7. 
\vskip 0.5cm
{\bf Table Caption}

Table I. Some relevant physical quantities for some representative values of 
$B$ and $\sigma$ using the standard prefactor.
\pagebreak

\pagebreak
\vskip 4cm
\begin{figure}[ht]
\vskip 15truecm
\includegraphics{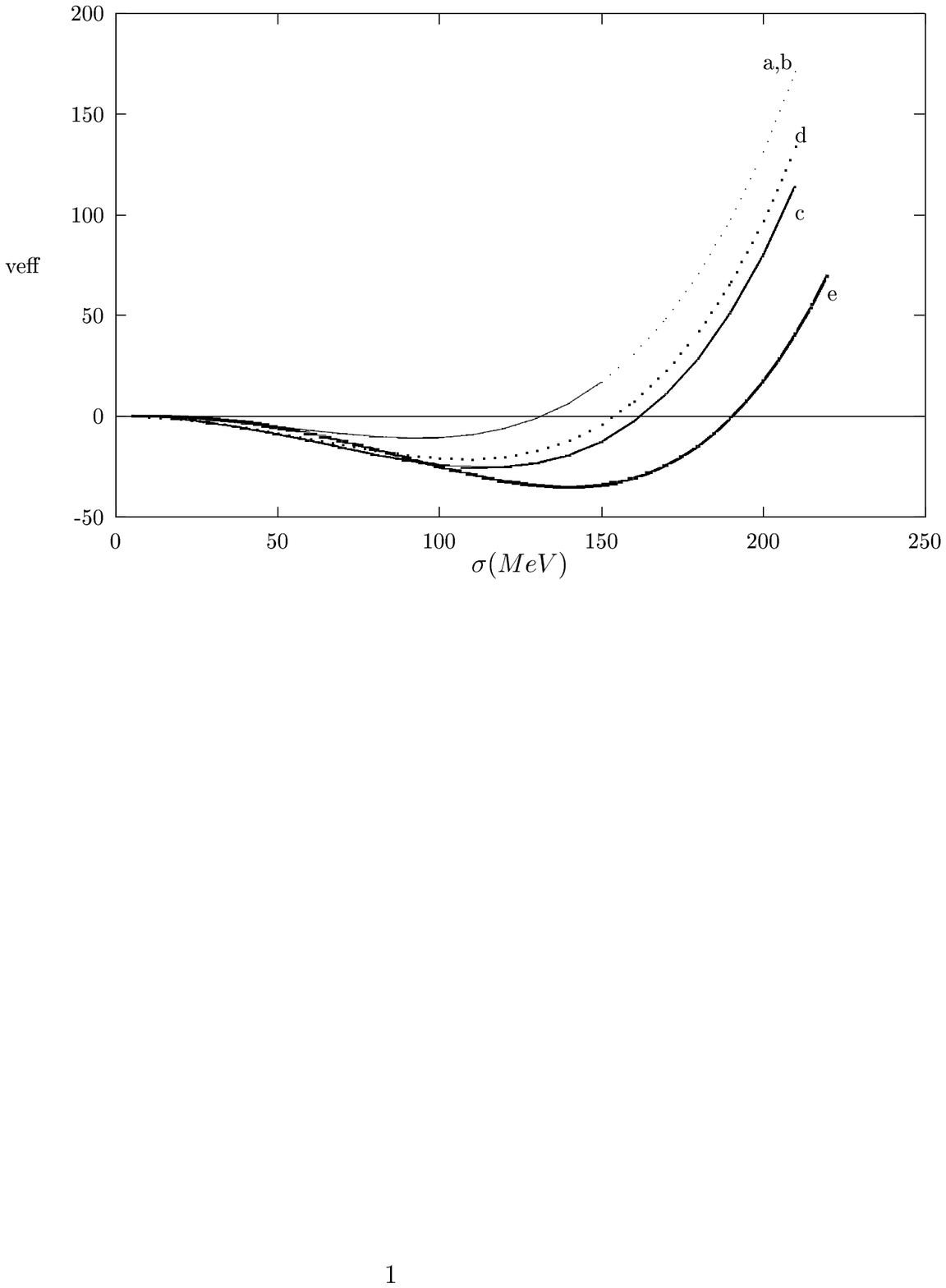}

\end{figure}
\pagebreak
\vskip 4cm
\begin{figure}[ht]
\vskip 15truecm
\includegraphics{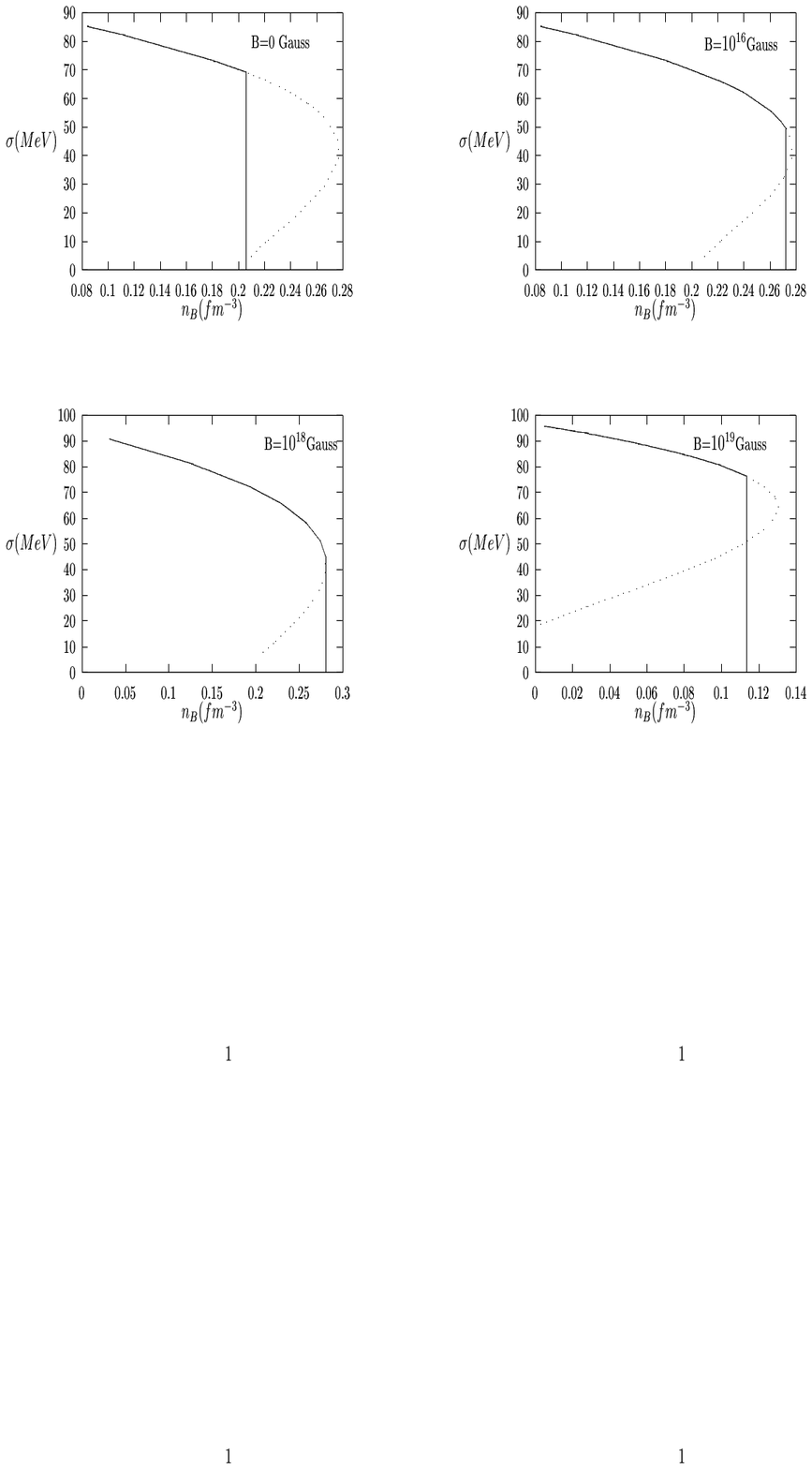}

\end{figure}
\pagebreak
\vskip 4cm
\begin{figure}[ht]
\vskip 15truecm
\includegraphics{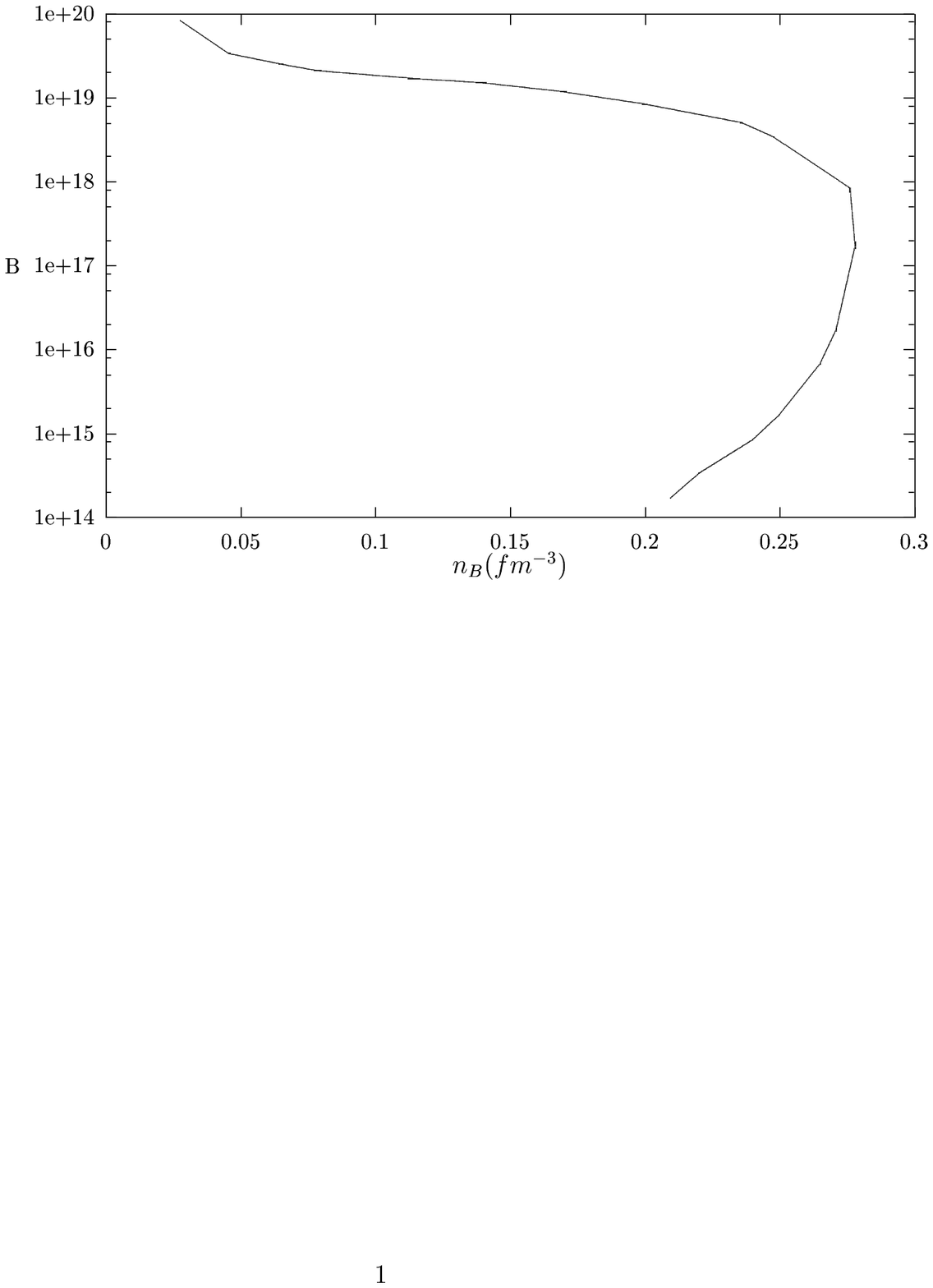}

\end{figure}
\pagebreak
\vskip 4cm
\begin{figure}[ht]
\vskip 15truecm
\includegraphics{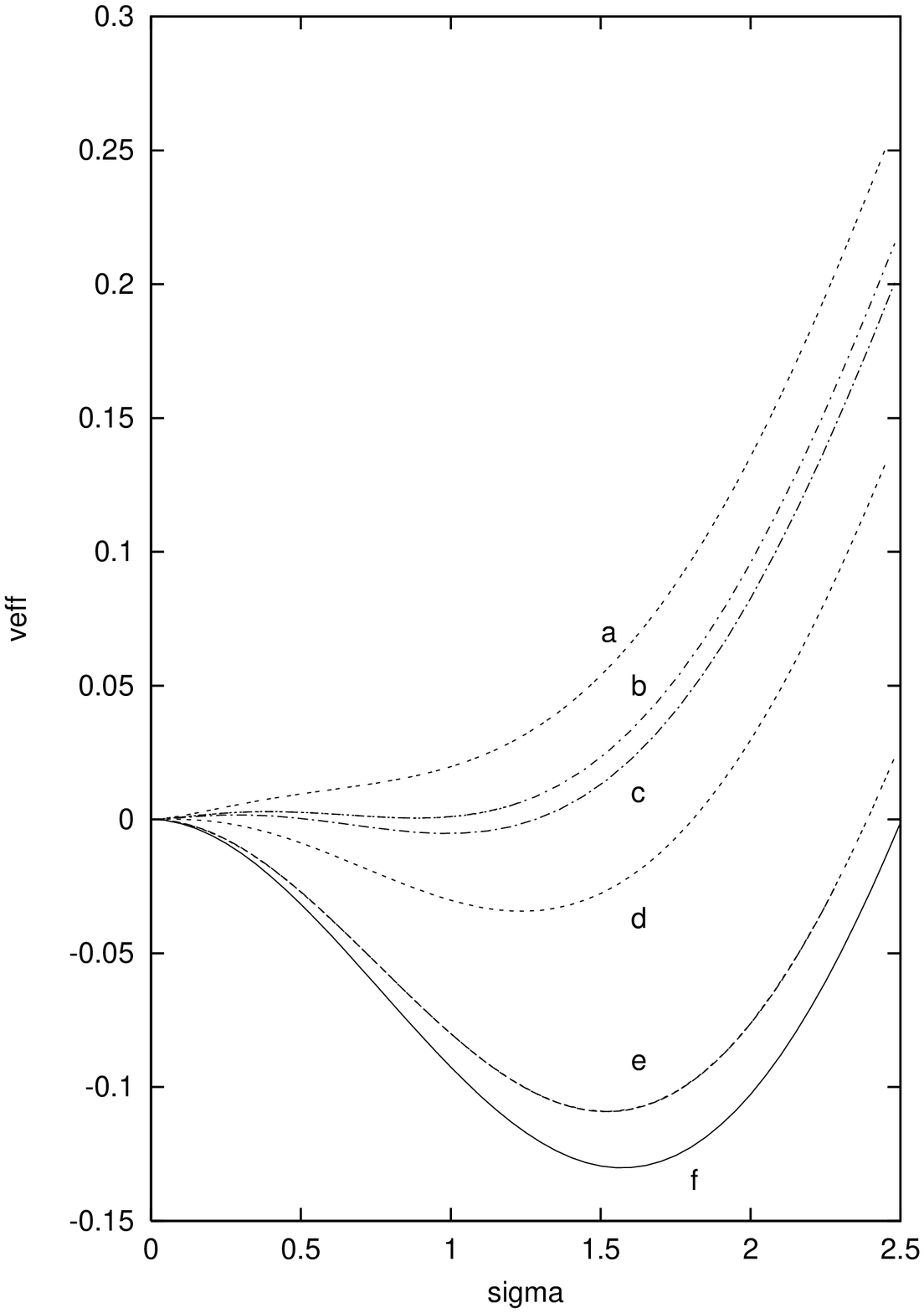}

\end{figure}
\pagebreak
\vskip 4cm
\begin{figure}[ht]
\vskip 15truecm
\includegraphics{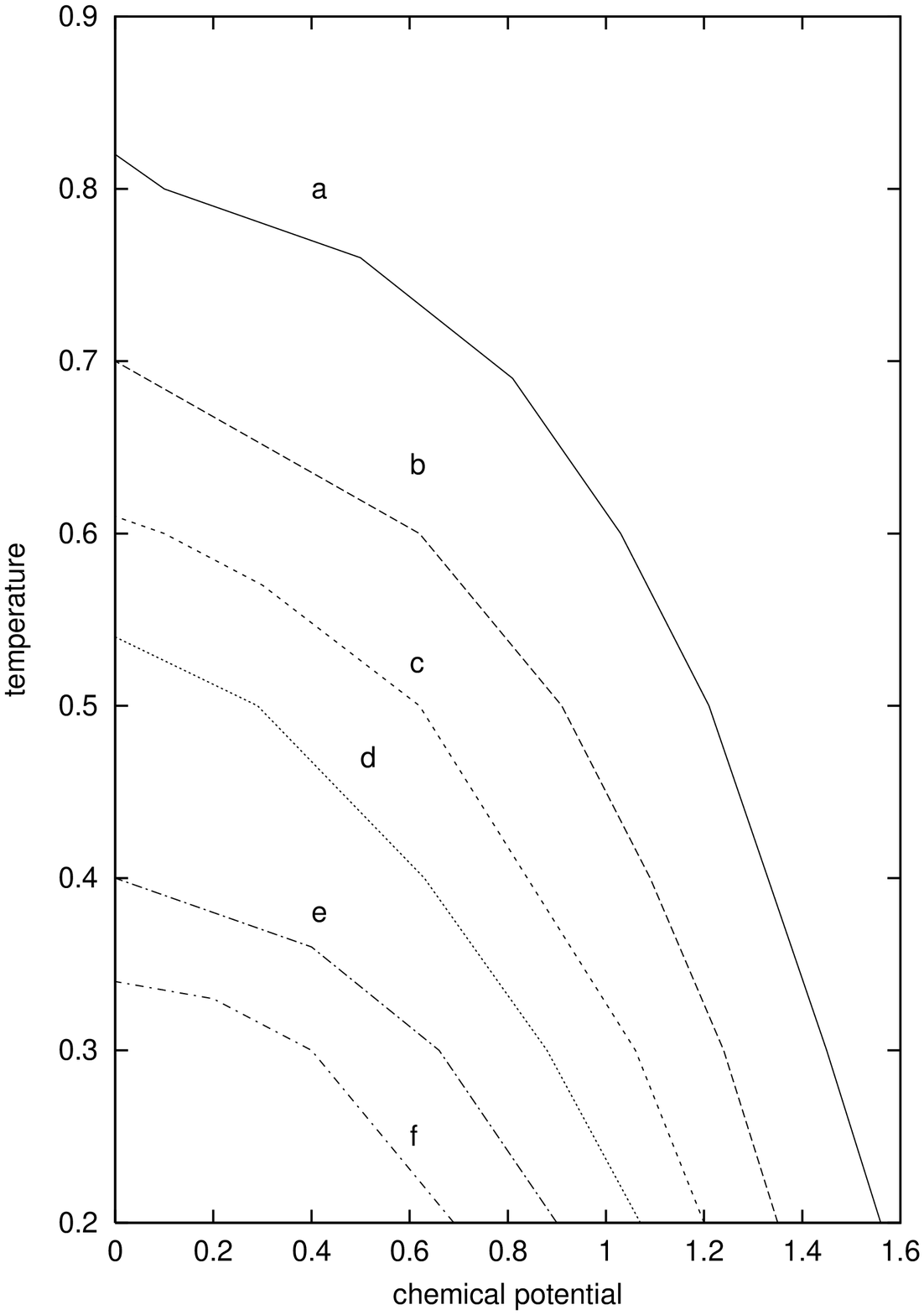}

\end{figure}
\pagebreak
\vskip 4cm
\begin{figure}[ht]
\vskip 15truecm
\includegraphics{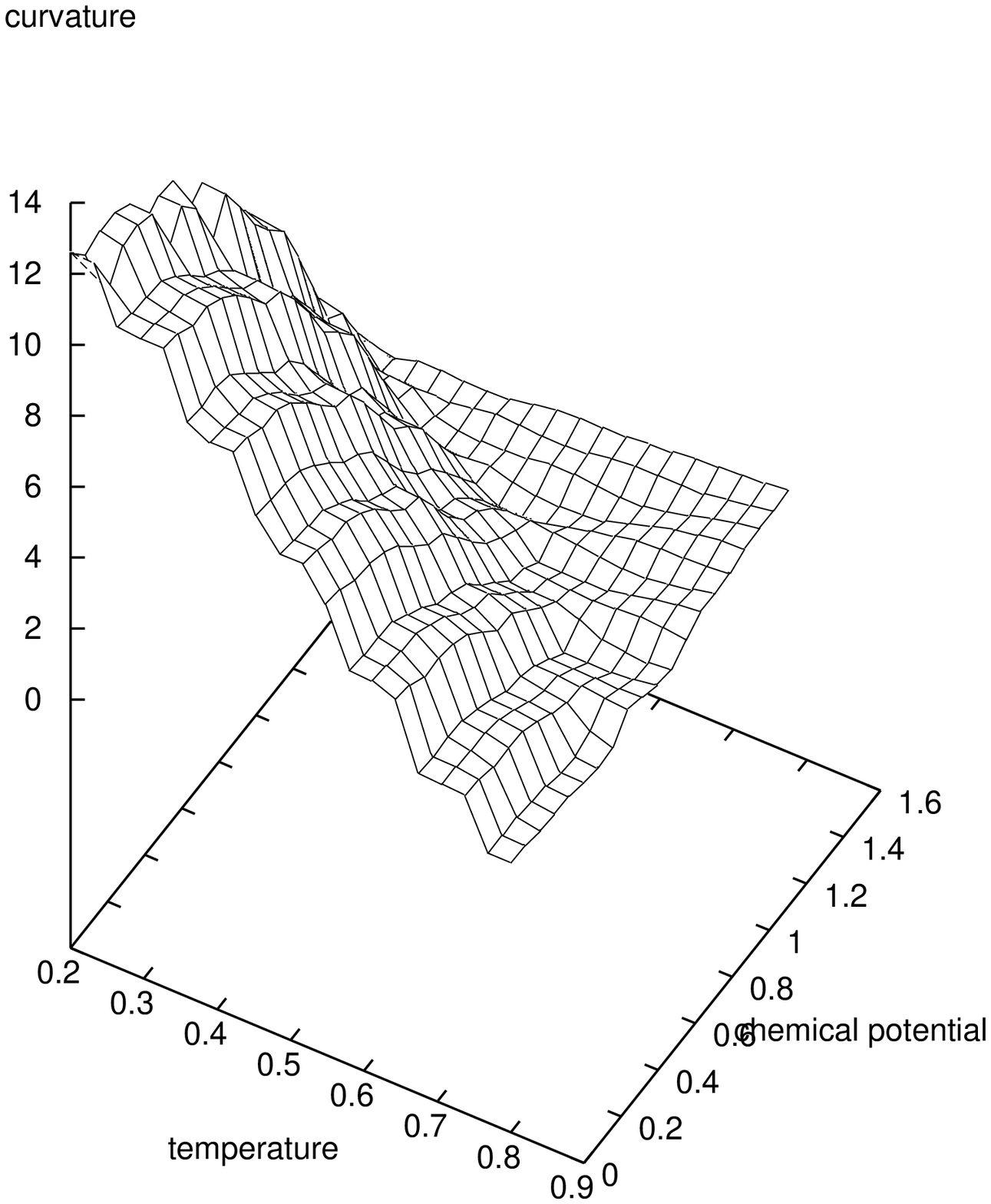}

\end{figure}
\pagebreak
\vskip 4cm
\begin{figure}[ht]
\vskip 15truecm
\includegraphics{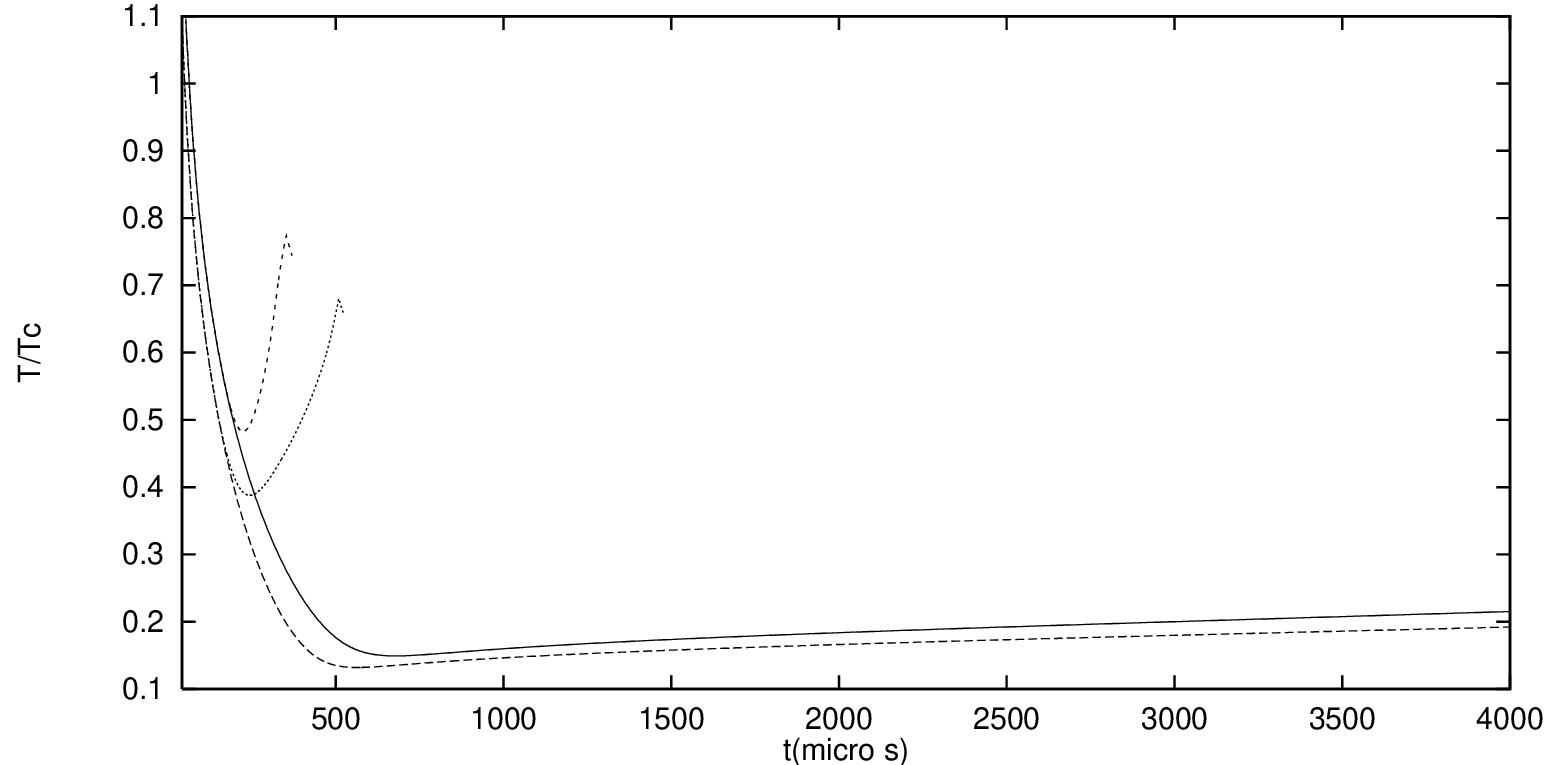}

\end{figure}
\pagebreak
\vskip 4cm
\begin{figure}[ht]
\vskip 15truecm
\includegraphics{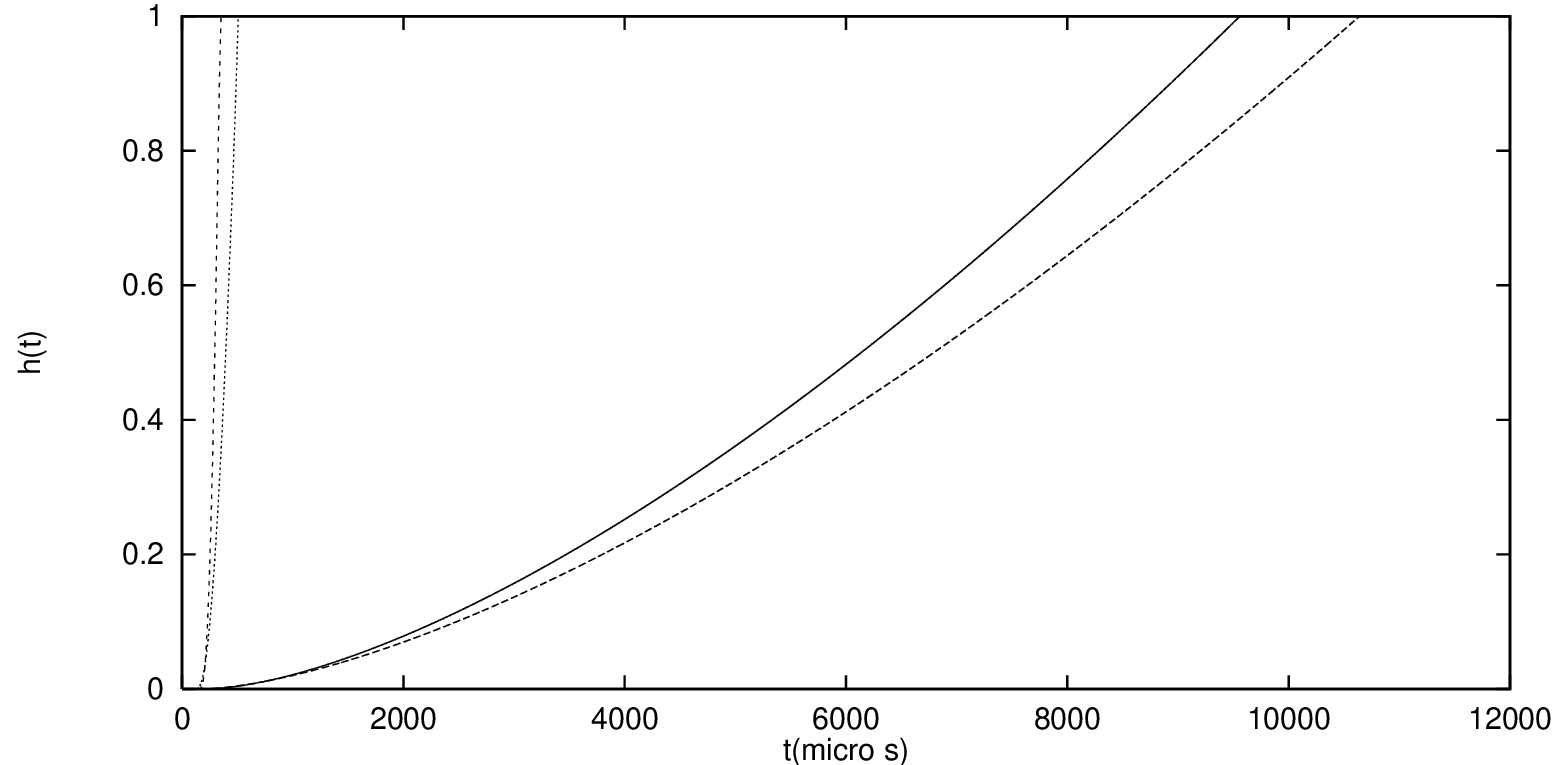}

\end{figure}
\end{document}